\documentclass[aps,prb,a4paper,superscriptaddress,twocolumn,showpacs,amsmath,amssymb]{revtex4} 
\usepackage{graphicx,epsfig} 
\allowdisplaybreaks
\begin{document} 

\newcommand{\s}{\sigma}
\newcommand{\up}{\uparrow}
\newcommand{\dw}{\downarrow}
\newcommand{\h}{\mathcal{H}}
\newcommand{\g}{\mathcal{G}^{-1}_0}
\newcommand{\D}{\mathcal{D}}
\newcommand{\A}{\mathcal{A}}
\newcommand{\K}{\textbf{k}}
\newcommand{\Q}{\textbf{q}}
\newcommand{\T}{\tau_{\ast}}
\newcommand{\io}{i\omega_n}
\newcommand{\eps}{\varepsilon}
\newcommand{\+}{\dag}
\newcommand{\su}{\uparrow}
\newcommand{\giu}{\downarrow}
\newcommand{\0}[1]{\textbf{#1}}
\newcommand{\ca}{c^{\phantom{\dagger}}}
\newcommand{\cc}{c^\dagger}
\newcommand{\da}{d^{\phantom{\dagger}}}
\newcommand{\dc}{d^\dagger}
\newcommand{\be}{\begin{equation}}
\newcommand{\ee}{\end{equation}}
\newcommand{\bea}{\begin{eqnarray}}
\newcommand{\eea}{\end{eqnarray}}
\newcommand{\ba}{\begin{eqnarray*}}
\newcommand{\ea}{\end{eqnarray*}}
\newcommand{\dagga}{{\phantom{\dagger}}}
\newcommand{\bR}{\mathbf{R}}
\newcommand{\bQ}{\mathbf{Q}}
\newcommand{\bq}{\mathbf{q}}
\newcommand{\bqp}{\mathbf{q'}}
\newcommand{\bk}{\mathbf{k}}
\newcommand{\bh}{\mathbf{h}}
\newcommand{\bkp}{\mathbf{k'}}
\newcommand{\bp}{\mathbf{p}}
\newcommand{\bRp}{\mathbf{R'}}
\newcommand{\bx}{\mathbf{x}}
\newcommand{\by}{\mathbf{y}}
\newcommand{\bz}{\mathbf{z}}
\newcommand{\br}{\mathbf{r}}
\newcommand{\Ima}{{\Im m}}
\newcommand{\Rea}{{\Re e}}

\title{Strongly Correlated Superconductivity rising from a Pseudo-gap Metal}
\author{Marco Schir\'o}
\affiliation{International School for Advanced Studies (SISSA), and CRS Democritos, CNR-INFM,
Via Beirut 2-4, I-34014 Trieste, Italy} 
\author{Massimo Capone}
\affiliation{SMC, CNR-INFM, and Universit\`a di Roma ``La Sapienza'', 
Piazzale Aldo Moro 2, I-00185 Roma, Italy} 
\affiliation{Istituto dei Sistemi Complessi, CNR, Via dei Taurini 19, I-00185 Roma, Italy}
\author{Michele Fabrizio} 
\affiliation{International School for Advanced Studies (SISSA), and CRS Democritos, CNR-INFM,
Via Beirut 2-4, I-34014 Trieste, Italy}
\affiliation{The Abdus Salam International Centre for Theoretical Physics 
(ICTP), P.O.Box 586, I-34014 Trieste, Italy} 
\author{Claudio Castellani}
\affiliation{SMC, CNR-INFM, and Universit\`a di Roma ``La Sapienza'', 
Piazzale Aldo Moro 2, I-00185 Roma, Italy}

\date{\today} 
\pacs{74.20.Mn, 71.27.+a, 71.30.+h, 71.10.Hf}
\begin{abstract}

We solve by Dynamical Mean Field Theory a toy-model which has a phase diagram strikingly similar to that of high $T_c$ superconductors: 
a bell-shaped superconducting region adjacent the Mott insulator and a normal phase that evolves from a conventional Fermi liquid 
to a pseudogapped semi-metal as the Mott transition is approached.    
Guided by the physics of the impurity model that is self-consistently solved within Dynamical Mean Field Theory, we introduce an 
analytical ansatz to model the dynamical behavior across the various phases which fits very accurately the numerical data. The ansatz is based on the assumption that the wave-function renormalization, that is very severe especially in the pseudogap phase close to the Mott transition, is perfectly canceled  by the vertex corrections in the Cooper pairing channel.
A remarkable outcome is that a superconducting state can develop even from a pseudogapped normal state, in which there are no low-energy quasiparticles. The overall physical scenario that emerges, although unraveled in a specific model and in an infinite-coordination Bethe lattice, can be interpreted in terms of so general arguments to suggest that it can be realized in  other correlated systems.    

\end{abstract}

\maketitle

\section{Introduction}

How can high-temperature superconductivity emerge out of a pseudogap metal is one of the standing puzzles posed by the cuprate 
superconductors. Indeed one would expect that a pseudogapped metal is not ideal for superconductivity mainly  
for two reasons: the pseudogap reduces the density of states at the Fermi level and, more worrisome, it is likely 
to cut off the BCS singularity in the Cooper channel. As a consequence, a very strong pairing should be required to turn such an 
unconventional metal into a high-$T_c$ superconductor. 
Many alternative proposals have been put forward to reconcile the existence of a pseudogap in the underdoped normal phase, 
which appears at a temperature $T_*$, with the occurrence of superconductivity (SC) below a critical temperature $T_c$ 
that may become significantly smaller than $T_*$ for deeply underdoped systems. 

The simplest explanation is to associate the opening of a pseudogap to the existence of preformed Cooper pairs 
well above $T_c$\cite{Randeria-xover,Haussmann-xover,Strinati-xover1,Tremblay-xover,Tsvelik-PG}.
This is compatible with the low dimensionality and the high critical temperature of the cuprates, 
which cooperate to enhance phase fluctuations of the order parameter, leading to a wide region where pairs are already formed 
yet lacking phase coherence. This scenario is experimentally supported by the strong diamagnetic response observed in 
an extended region above $T_c$\cite{Ong-PRB}. However, even though this point of view is certainly reasonable close to $T_c$, 
its application close to the pseudogap temperature scale $T_*$ and for small doping $x$ is definitely more questionable.
It is indeed well established that while $T_*$ increases monotonically as the doping $x\to 0$, both $T_c$ and the superfluid density 
vanish. Within the preformed-pair picture, this would correspond to very strong coupling leading to localized pairs in real space, 
which are hardly reconciled with the momentum-space nodal structure of the pseudogap observed by angle-resolved 
photoemission.~\cite{Campuzano-arc} 

A different point of view interprets the pseudogap as due primarily to a competing ordered 
phase\cite{Varma-QCP,Castellani-QCP,Tallon&Loram-QCP,Millis-QCP,Coleman-QCP,Si-QCP} or arising from fluctuating competing orders 
among which $d$-wave superconductivity prevails below $T_c$ (see Ref.~\onlinecite{Lee&Nagaosa&Wen-RMP} and references therein). 
Although compatible with many experimental evidences, these proposals pose, in our opinion, several theoretical questions. 
For instance, even if we assume that the pseudogap phase is a fluctuating mixture of 
different orders, we are still left with the question about the nature of the underneath normal phase unstable to all the above 
competing orders. A common belief is that the antiferromagnetic ground state of the undoped parent compounds~\cite{Zhang-SU(5)} 
or another state very close in energy~\cite{PWA} must be interpreted as the {\sl ancestor} phases that naturally evolve upon doping 
into a novel state of matter~\cite{Lee&Nagaosa&Wen-RMP} - a fluctuating mixture of pseudogap phases - rather than into a 
{\it bona fide} normal metal that, below $T_c$, turns superconducting. 

In this work we do not intend to enter these controversial issues in the context of cuprates. Rather, we want to unravel in all 
its facets a similar phenomenology - pseudogap normal phase which turns into a {\sl high-temperature} superconductor - 
that we recently discovered by solving with Dynamical Mean Field Theory (DMFT)~\cite{DMFT} a two-orbital Hubbard model inspired by 
fullerene superconductors.~\cite{Capone-PRL-2} We think that providing an exhaustive analysis of the pseudogap phenomenon in a 
strongly correlated model that can be exactly solved, {\it albeit} in an infinite-coordination lattice, may shed light on more 
realistic models for the cuprates, which are harder to deal with, both analytically and numerically.
We are going to show that in this two-orbital model, akin to models for cuprates, superconductivity is the low-temperature winner 
among competing phases. In our case, two of these competing phases are homogeneous and symmetry-invariant: a conventional 
Fermi-liquid metal and an intrinsic pseudogap phase, i.e., a single quantum phase with zero entropy at $T=0$.
The connection between the superconducting and the intrinsic pseudogap phases is intriguing, as the latter proves to be a fertile 
ground for superconductivity.
The analysis of the self-energy in the superconducting phase reveals that the onset of superconductivity gives rise to much more 
coherent excitations than in the normal pseudogap state, where there are no quasiparticles and the self-energy diverges 
at low frequency.
The way in which this singularity is regularized in the superconducting phase bears strong similarities with the effect of 
non magnetic impurities in an $s$-wave superconductor, where the processes leading to anomalies in the normal phase do not cut-off 
the superconducting instability, or, in other words, they are {\it non-pairbreaking}. Despite the regularization of the low-frequency 
anomalies, where the superconducting gap develops, our correlated superconductor still presents the pseudogap energy scale 
at high-frequency, and these two scales have opposite behavior as the doping goes to zero.

\begin{figure}
\includegraphics[width=6cm]{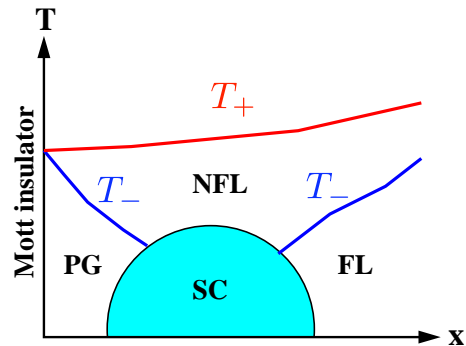}
\caption{\label{phd-rough}(Color online) Sketch of the phase diagram of the model Eq.~(\ref{Ham}). $x$ is a parameter that measures 
the deviation from the Mott insulator.  FL, NFL, SC and PG stand for normal Fermi-liquid, non-Fermi-liquid, 
superconducting and normal pseudogap phases, respectively. $T_-$ and $T_+$ are cross-over lines, see the text.}
\end{figure}

Although we are going to review quite in detail the model and its properties in the following sections, 
we think is worth anticipating here some of the main results.  
In Fig.~\ref{phd-rough} we sketch the phase diagram of the model.~\cite{Capone-PRL-2}. 
On the x-axis we plot a parameter that controls the distance from a Mott insulating phase like doping or pressure. 
The acronyms FL, NFL, SC and PG stand for normal Fermi-liquid, non-Fermi-liquid, superconducting and normal 
pseudogap phases, respectively. The temperature scale $T_-$ (two branches) identifies the crossovers from 
the FL and PG phases into the almost critical NFL region. 
The latter is not pseudogapped and it is characterized by incoherent single-particle excitations with dispersion
and inverse-lifetime controlled by a single scale $T_+$. The Mott transition occurs when $T_+=T_-$ on the 
pseudogap side of the superconducting region. 

The phase diagram Fig.~\ref{phd-rough} bears striking similarities to that of cuprates, identifying the $T_-$ branch on the
``underdoped'' side with $T_*$. From a purely theoretical point of view, the diagram closely resembles the quantum-critical-point (QCP) 
scenario proposed in many contexts, including cuprates~\cite{Varma-QCP,Castellani-QCP,Tallon&Loram-QCP} 
and heavy-fermion compounds~\cite{Millis-QCP,Coleman-QCP,Si-QCP}, as well as the phase diagram of the $t-J$ model 
for cuprates~\cite{Kotliar&Liu,Suzumura-RVB,Lee&Nagaosa&Wen-RMP} within the Resonating Valence Bond (RVB) framework.~\cite{PWA}
We postpone a critical comparison with those to the ending section of the paper. 

The paper is organized as follows. In Section II we introduce our model and approach. Sec. III is devoted to the 
DMFT phase diagram, while
Sec. IV summarizes the information that we can obtain from the impurity model which corresponds to our lattice model through DMFT. 
Secs. V and VI introduce modelizations of the self-energy in the normal and superconducting phases, respectively. 
Finally, Sec. VII is dedicated to conclusions with particular attention to the relation between our 
results and theoretical proposals for the cuprates.

\section{The model}
\label{Sec:Model}

The model that we study is a two-orbital Hubbard model with the Hamiltonian 
\bea
\mathcal{H} &=& -\sum_{ij}\,\sum_{a=1}^2\,\sum_\sigma\, t_{ij}\,\Big(c^\dagger_{i,a\s} c^\dagga_{j,a\s} + H.c. \Big) 
\nonumber\\ 
&& + \frac{U}{2}\sum_i\,\left(n_i-2\right)^2 \nonumber \\
&& -2J\, \sum_i\, \Big( T_{i,x}^2 + T_{i,y}^2\Big),\label{Ham}
\eea 
where $c^\dagger_{i,a\sigma}$ and $c^\dagga_{i,a\s}$ creates and annihilates, respectively, one electron at site $i$ 
in orbital $a=1,2$ with spin $\sigma$,  $n_i = n_{i,1}+n_{i,2}$ is the on-site occupation
number, being $n_{i,a} = \sum_{\sigma} c^\dagger_{i,a\sigma} 
c^\dagga_{i,a\sigma}$, and finally, 
\be
T_{i,\alpha} = \frac{1}{2} \sum_{a,b=1}^2\,\sum_\sigma\, c^\dagger_{i,a\sigma}\, 
\tau^\alpha_{ab}\, 
c^{\phantom{\dagger}}_{i,b\sigma} 
\ee
are orbital pseudo-spin-1/2 operators, $\tau^\alpha$ ($\alpha=x,y,z$) being the Pauli matrices. 
We assume hereafter that the exchange constant $J$ is positive, hence favors low-spin atomic configurations.
This model was introduced in Ref.~\onlinecite{Capone-PRL-2} to 
mimic an $e\otimes E$ Jahn-Teller coupling to a local doubly-degenerate phonon mode that prevails over 
the conventional Coulomb exchange. In this case, the Jahn-Teller coupling leads effectively to inverted Hund's rules, on provision that 
the phonon frequency is high enough to neglect retardation effects. The original purpose was to study a simplified 
model that shared the same physics of alkali-doped fullerene superconductors, where pairing is mediated by 
eight fivefold-degenerate local vibrational modes, $t\otimes H$ Jahn-Teller coupled to the threefold degenerate LUMO 
of C$_{60}$.~\cite{Gunnarsson-RMP,Capone-PRL-1,Capone-Science} Apart from a constant term, the Hamiltonian 
(\ref{Ham}) can be alternatively written as  
\bea
\mathcal{H} &=& -\sum_{ij}\,\sum_{a=1}^2\,\sum_\sigma\, t_{ij}\,\Big(c^\dagger_{i,a\s} c^\dagga_{j,a\s} + H.c. \Big) \nonumber \\
&& + 
\frac{U}{2}\sum_i\,\sum_{a=1}^2\, \left(n_{i,a}-1\right)^2\nonumber\\
&& +\sum_i\, J_\perp\,\mathbf{S}_{i,1}\cdot\mathbf{S}_{i,2} + V\,n_{i,1}\,n_{i,2},\label{Ham-bis} 
\eea
with $J_\perp=4J$ and $V=U+J$, that also describes two Hubbard models labeled by the orbital index $a=1,2$,
coupled by an antiferromagnetic exchange $J_\perp$ and by a strong repulsion $V$. 

The interaction term proportional to $J$ in (\ref{Ham}) is easily shown to generate an attraction that leads to 
an order parameter of s-wave symmetry associated to the operator
\be
\Delta_\bk = c^\dagger_{\bk,1\su}c^\dagger_{-\bk,2\giu} + c^\dagger_{-\bk,2\su}c^\dagger_{\bk,1\giu}.
\label{SC-chan}
\ee
The bare scattering amplitude in this channel is $A=-2J$, and it would induce a superconducting instability in the absence of 
Coulomb repulsion. The introduction of $U$ weakens the attraction, leading to $A = -2J+U$. Therefore, at least in weak coupling, 
$U,J\ll W$ ($W$ being the non-interacting bandwidth), one expects the model to describe a BCS superconductor for $U\leq 2J$, 
and a normal metal for larger repulsion. In principle, if the non-interacting Fermi surface accidentally has nesting, other 
weak-coupling instabilities like, e.g., magnetism can occur.
In our calculation we did not consider such commensurate phases, because they are related to specific aspects of the lattice, 
and we want to focus on more basic and general properties. We note that the competition between attraction and repulsion, 
namely between $J_\perp$ and $V$ in (\ref{Ham-bis}), offers the opportunity to investigate how $s$-wave superconductivity can emerge 
at all in spite of strong correlations. This issue is not commonly touched in the context of cuprates, where the emphasis is mostly put 
on the appearance of $d$-wave superconductivity (whose order parameter lives on the bonds) in the presence of purely on-site repulsion, 
while longer-range contributions to the interaction are often ignored.~\cite{Plekhanov-PRL} 

Let us move now to the opposite limit of strong repulsion, $U\gg W$, and consider the half-filled case. 
In this limit the model is a Mott insulator, each site being occupied by two electrons that can not move coherently. 
When $J=0$ the model maps onto an SU(4) Heisenberg model. If the hopping is restricted to nearest neighbor, the ground state 
of this model is dimerized in one dimension~\cite{Assaraf-SU(4)}, while in a two dimensional square lattice is still unclear 
whether is spin-liquid or N\'eel ordered.~\cite{Assaad-SU(N),Paramekanti-2006} When the attraction is switched on and $J\gg W^2/U>0$ 
the attractive interaction prevails and two electrons on each site $i$ lock into the singlet state 
\be
{\frac{1}{\sqrt{2}}}\,\left(c^\dagger_{i,1\su}c^\dagger_{i,2\giu}+c^\dagger_{i,2\su}c^\dagger_{i,1\giu}\right)\,|0\rangle,
\label{singlet}
\ee
stable to the weak inter-site superexchange. In this case, the Mott insulator is non-magnetic and translationally invariant - 
a local version of a valence-bond crystal -  regardless the structure of the hopping matrix element and  the dimensionality and 
topology of the lattice.
A sufficient degree of frustration, either due to the lattice topology or to the hopping amplitudes $t_{ij}$'s, 
disfavors a magnetic state and makes it possible for this non-magnetic phase to survive decreasing
$U/W$ down to the Mott transition, which is in turn pushed to a finite $U_c$ when frustration eliminates nesting.
When this happens, one should na\"{\i}vely expect, increasing $U/W$ at fixed $J/W\ll 1$, first a very narrow BCS superconducting 
region for $0\leq U\leq 2J$, followed by a normal metal that eventually gives way to a non-magnetic Mott insulator 
when $U\geq U_c \sim W$. Seemingly, doping the non-magnetic Mott insulator 
at $U> U_c\gg J$ should bring to a normal metal that, as doping increases, gets less and less correlated. 
   
This na\"{\i}ve expectation, based essentially on the value of the bare scattering amplitude $A=-2J+U$ in the 
singlet Cooper-channel Eq.~(\ref{SC-chan}), turns out to be wrong at least in the two cases that have been so far considered: (i) 
an infinite-coordination Bethe lattice~\cite{Capone-PRL-2}, which is exactly solved by DMFT; (ii) a one dimensional 
chain~\cite{Fabrizio-1D-PRL}. We will briefly mention the latter in the last Section, while in what follows we concentrate 
mostly on the DMFT results for the Bethe lattice. In the following we first recall the basic ideas behind 
DMFT~\cite{DMFT} and our implementation.

DMFT extends the idea of classical mean-field to the quantum domain: a lattice model is approximately solved by 
solving the quantum problem of a single site subject to 
a ``dynamical Weiss field'' that describes the action of the rest of the lattice sites on the given site 
(assumed to be equivalent to any other). As in classical mean-field, the mapping is exact only in infinite coordination-lattices.
The effective action of the local degrees of freedom reads  

\bea
\label{seff}
S_{eff}& = &\int_0^{\beta}d\tau\,d\tau^{\prime}\, c^{\dagger}_{0\alpha\sigma}(\tau)\,
{\cal{G}}_0^{-1}\left(\tau-\tau^{\prime}\right)^{\alpha\beta}_{\sigma\sigma^{\prime}}\, 
c_{0\beta\sigma^{\prime}}(\tau^{\prime})+\nonumber \\
&&+ S_0[c_{0\alpha\sigma},c^{\dagger}_{0\alpha\sigma}],
\eea
where ${\cal G}_0^{-1}$ is the Weiss field, $\alpha$ and $\beta$ are orbital indices and $\sigma$ and $\sigma^{\prime}$ are 
spin indices. $S_0$ is the local part of the action and contains all local interaction terms of the lattice Hamiltonian.
The mean-field scheme is closed by imposing between the Weiss field and the local Green's function 
$G\left(\tau-\tau'\right)$, computed by the action (\ref{seff}), a self-consistency relation that contains the information about 
the original lattice model through the non-interacting density of states (DOS), 
For a Bethe lattice with nearest neighbor hopping and bandwidth $W$, which we consider hereafter, the self-consistency reads
\be
\label{self}
{\cal G}_0^{-1}(i\omega_n)^{\alpha\beta}_{\sigma\sigma^{\prime}} = i\omega_n -E^{\alpha\beta}_{\sigma\sigma^{\prime}} -
\frac{W^2}{16}G(i\omega_n)^{\alpha\beta}_{\sigma\sigma^{\prime}},
\ee
where $E$ is the matrix of the single-particle terms of the on-site Hamiltonian (chemical potential, magnetic field, hybridization) 
and $G(i\omega_n)$ is the Fourier transform of the local Green's function $G(\tau)^{\alpha\beta}_{\sigma\sigma^{\prime}} = 
-\langle T_{\tau} c_{0\alpha\sigma}(\tau)c^{\dagger}_{0\beta\sigma^{\prime}}(0)\rangle_{S_{eff}}$. Eqs.~(\ref{seff}) and (\ref{self}) 
can be viewed as a set of two coupled equations for $G$ and ${\cal G}_0^{-1}$. 
In practice, one needs to solve (\ref{seff}) for a given choice of ${\cal G}^{-1}_0$ and obtaining $G$. Using (\ref{self}) 
one finds a new value of ${\cal G}^{-1}_0$. The procedure is iterated until convergence is achieved. It is evident that the 
computation of $G$ is the hard part of the calculation. 
An important observation is that the effective local theory can be represented as an impurity model whose hybridization function 
coincides with the dynamical Weiss field. In practice, a DMFT calculation amounts to solve an Anderson impurity model iteratively in 
order to self-consistently determine its hybridization function.

The solution of the impurity model requires either an approximate numerical method, or an ``exact'' numerical approach. In this work 
we use exact diagonalizaton at $T=0$\cite{krauth}, which amounts to approximate the continuous bath of the impurity model by 
a discrete set of energy levels hybridized with the impurity. The method has been shown to converge exponentially as function of 
the number $N_b$ of bath levels. For example, $N_b = 5$ already gives quite reliable results for the phase diagram and  
thermodynamic observables of a single band Hubbard model. 
In this work we will use 6 bath levels for each orbital, which gives $N_b=12$ in total. An important aspect of the exact 
diagonalization method is the way in which the continuous bath is approximated. This is implemented by minimizing a suitably 
chosen distance between two functions. The distance is typically computed on the imaginary axis on a Matsubara grid corresponding 
to an effective temperature $\tilde\beta$, which is not to be confused with the physical temperature, always set to $T=0$. 
In this work we typically use $\tilde\beta =400/W$, and define a distance that weights more low frequencies\cite{capone-1d}. 
Specifically, we minimize
\be
\chi = \sum_{n}\frac{\vert{\cal G}_0^{-1}-({\cal G}_0^{-1})_{discrete}\vert}
{\omega_n}
\ee
$({\cal G}_0^{-1})_{discrete}$ being the discrete version of the Weiss field, with the sum extending up to a maximum frequency 
of order $4U$. 

Finally, in order to describe the superconducting phase, the bath includes superconducting terms leading to a Weiss field with 
anomalous components. Consequently the impurity Green's function has also an anomalous term 
$F_{\alpha\beta}(\tau)=- \langle T_{\tau} c_{\alpha\uparrow} (\tau) c_{\beta\downarrow}(0)\rangle$, which  is used together with $G$ 
to build the matrix $\hat{G}(i\omega_n)$ in the Nambu-Gor'kov formalism.
Seemingly one define a matrix Weiss field with diagonal, ${\cal G}^0(i\omega_n)^{-1}$, and off-diagonal, ${\cal F}^0(i\omega_n)^{-1}$, 
components. The self-consistency condition (\ref{self}) can be rewritten in the same formalism straightforwardly.

\section{The phase diagram in a Bethe lattice}

\begin{figure}
\includegraphics[width=8cm]{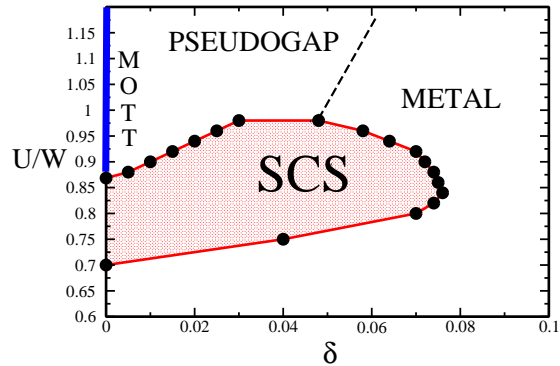}
\caption{\label{phd} (Color online) DMFT phase diagram of the model Eq.~(\ref{Ham}) as function of $U/W$ and doping $\delta$. 
SCS stands for strong-correlated superconductivity.}
\end{figure} 

In Fig.~\ref{phd} we show the DMFT phase diagram of model (\ref{Ham}) at fixed $J=0.05\,W$ in a Bethe lattice as function of doping 
and $U/W$ around the Mott transition, $U_c\simeq 0.87\,W$.~\cite{Capone-PRL-2}. The first remarkable thing to note is the appearance 
of superconductivity, denoted as SCS (for Strongly Correlated Superconductivity, see below) in the figure, just around the Mott 
transition. The symmetry of the order parameter is that of Eq. (\ref{SC-chan}). 
Superconductivity even extends for $U\agt U_c$ within a finite doping interval and is preceded by a pseudogap metal 
and followed, at larger doping, by a more conventional normal metal, see Fig.~\ref{phd}. 
We emphasize that the discreteness of the spectra obtained in ED does not allow to unambiguously identify the pseudogap region, 
yet the results clearly show the evolution from one kind of normal state into the other, as already shown in 
Ref. \onlinecite{Capone-PRL-2}.

\begin{figure}[htb]
\includegraphics[width=8.7cm,height=6cm]{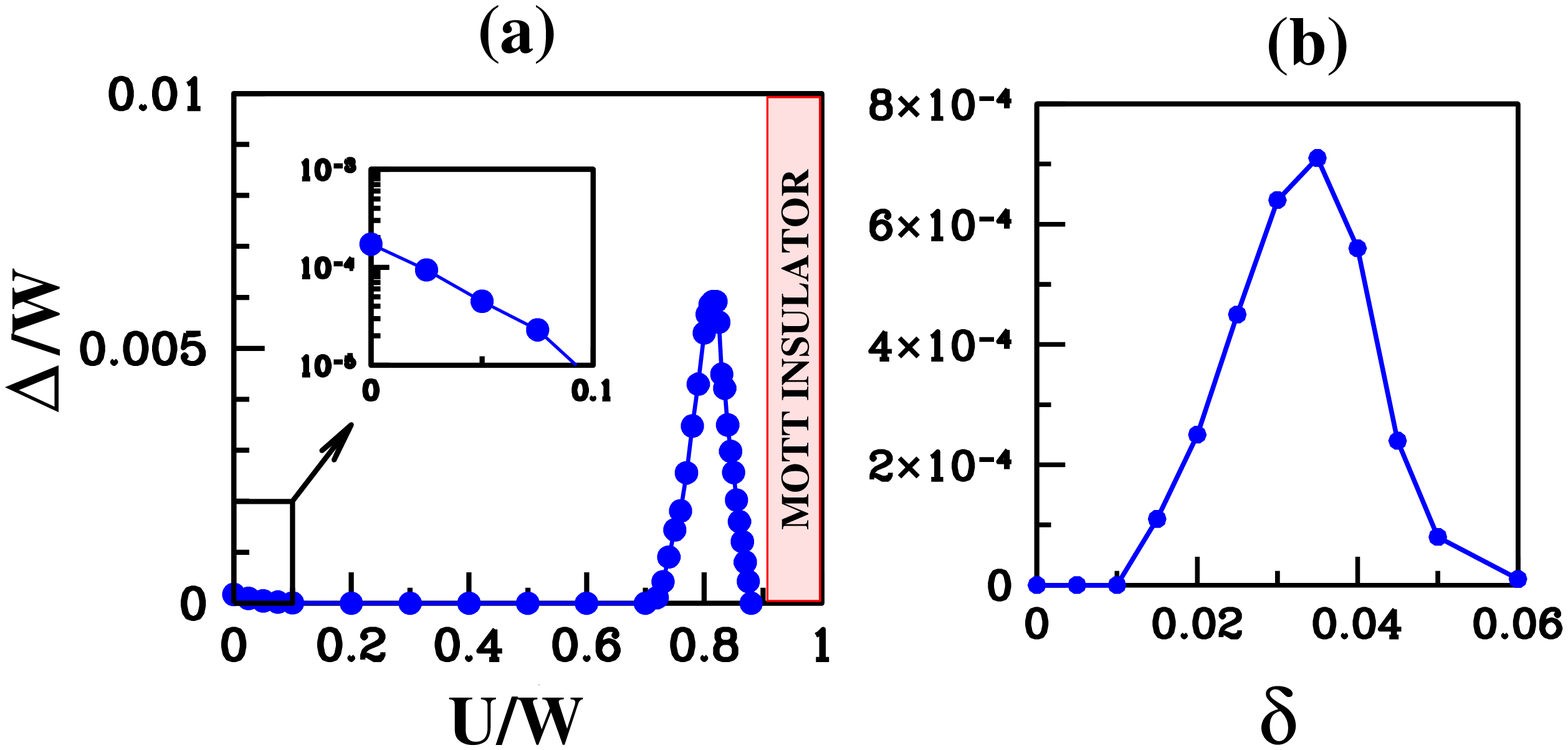}
\caption{\label{gap} (Color online) Superconducting gap at half-filling as function of $U/W$, panel (a), and for $U=0.92~W$ 
as function of doping away from half-filling, panel(b). The inset in panel (a) shows the gap at weak $U$ on an enlarged scale.}
\end{figure}

The strength of pairing is also surprising. In Fig.~\ref{gap} we draw the values of the  
superconducting gap $\Delta$ at half-filling as function of $U/W$, panel (a), and at $U=0.92~W$ as function of doping, panel (b). 
In panel (a) we also show on a smaller scale the same quantity in the BCS-like region for small $U\leq 2J$. The latter is 
exponentially small, in agreement with the BCS estimate $\Delta\sim 0.5 W\,\exp\left(-\pi W/8 J\right)\sim 2\times 10^{-4}W$, 
almost two orders of magnitude smaller that the values attained in the superconducting region around the Mott insulator. 

\begin{figure}[hbt]
\includegraphics[width=8.7cm,height=6cm]{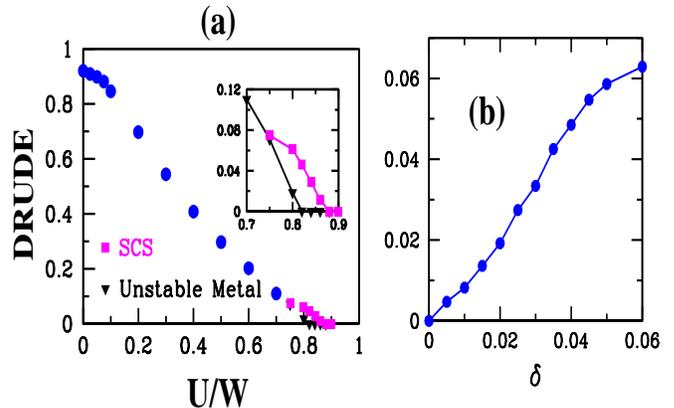}
\caption{\label{drude} (Color online) (a) Drude weight at half-filling as function of $U/W$. In the SCS region is also shown 
the Drude weight of the metastable normal solution, zoomed in the inset; (b) Drude weight at $U=0.92~W$ 
as function of doping.}
\end{figure}

This superconducting phase that re-emerges with strengthened pairing just before the Mott transition was named in 
Ref.~\onlinecite{Capone-Science} {\sl Strongly Correlated Superconductivity} (SCS) to emphasize its peculiar properties 
with respect to conventional BCS superconductors. Indeed, besides the large value of the gap in spite of  
the tiny attraction $J=0.05~W$, other features characterize SCS.   
In Fig.~\ref{drude}(a) we compare the values at half-filling of the Drude weight (zero-frequency contribution to the optical 
conductivity $\sigma(\omega)$) in the SCS phase and in what could be regarded as the {\sl normal phase}, namely the metastable 
solution obtained within DMFT by preventing gauge symmetry breaking. We note that the onset of superconductivity is accompanied by 
an increase of Drude weight, unlike what happens in a BCS superconductor. Remarkably, while the SCS Drude weight vanishes only at the 
Mott transition, the weight of the normal solution vanishes for a smaller $U$ when the zero-frequency spectral weight goes to zero, 
suggesting the opening of the pseudogap. This semiconducting normal-phase is metastable at half-filling, where the stable 
zero-temperature phase is superconducting, but it is stabilized away from half-filling as shown in the phase diagram of Fig.~\ref{phd}. 
Here, doping seems to have the same effect as in semiconductors, leading to a Drude weight linearly increasing with the number of 
holes, reported in panel (b) of  Fig.~\ref{drude}. 
The intrusion of SCS leads to a more pronounced increase of Drude weight that smoothly connects to the normal metal appearing for 
larger doping.

\subsection{Fermi liquid description}

Since the stable metallic phases at small $U/W$ at half-filling, or at large doping away from half-filling, do not show any particular 
anomaly, a Landau Fermi-liquid scenario is expected to be applicable to understand how superconductivity can emerge in spite of 
the fact that the bare scattering amplitude in the Cooper channel is repulsive. The Fermi-liquid behavior is indeed confirmed by 
the regular self-energies calculated by DMFT. Within perturbation theory, the effect of interaction on the low-energy 
single-particle properties can be absorbed into the so-called wave-function renormalization or quasiparticle residue $Z$ defined by
\be
\label{defZ}
\frac{1}{Z} = 1 - \left(\frac{\partial \Sigma(i\omega_n,k_F)}{\partial i\omega_n}\right)_{\omega_n\to 0},
\ee
where $\Sigma(i\omega_n,k_F)$ is the single-particle self-energy in Matsubara frequencies at the Fermi momentum, 
and by the reduction of the quasi-particle bandwidth $W\to W_*<W$, 
\be
\frac{W_*}{W} = Z\,\left[1 + \frac{1}{v_F^0}\left(\frac{\partial \Sigma(0,k)}{\partial k}\right)_{k=k_F}\right],
\ee
where $v_F^0$ is the bare Fermi velocity. $Z$ can be regarded as the component of the total single-particle spectral weight that 
is carried by coherent quasiparticle excitations. In general $Z\leq W_*/W$, the equality holding only in infinite-coordination 
lattices where the self-energy is momentum independent, $\Sigma(i\omega_n,k) \equiv \Sigma(i\omega_n)$.~\cite{Metzner&Vollardt} 

\begin{figure}[htb]
\includegraphics[width=6cm]{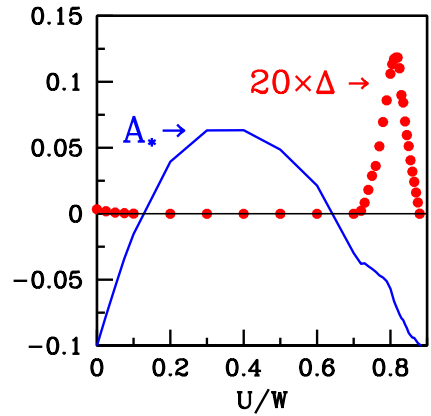}
\caption{\label{A*} (Color online) Ansatz for the scattering amplitude $A_*=Z\,U-2J$ versus $U/W$ using $Z$ extracted from a normal DMFT solution. 
Also shown is the superconducting gap 
multiplied by a factor 20 for graphic convenience.}
\end{figure}

Within Landau theory, considering a generic scattering channel with bare amplitude $\Lambda$, the renormalized value can be written as
\be
\Lambda_* = Z^2\, \Gamma_{\Lambda}\, \Lambda,
\ee
where $\Gamma_{\Lambda}$ includes the so-called vertex corrections. Approaching an interaction-driven Mott transition, $U\to U_c$, 
the quasiparticle residue vanishes $Z\propto U_c-U \to 0$, but the behavior of ${\Lambda}_*$ in different channels can be totally 
different due to a different relevance of vertex corrections.
Physical intuition suggests that the proximity to a Mott transition affects primarily charge fluctuations, that are severely 
suppressed, but it does not equally influence the way in which the charge is distributed between different spin and orbital states. 
Indeed, the localization of the charge leads to the formation of local moments and reflects an enhancement of the spin and orbital 
responses. This suggests that, while vertex corrections in the spin- and orbital-density channels can compensate the vanishing $Z$, 
the same cancellation does not take place in the charge-density channel. 
Following this observation, we argued in Ref.~\onlinecite{Capone-Science} that the on-site repulsion $U$ and exchange $J$ undergo 
different renormalization as the Mott insulator is approached. 
In particular, since the exchange term $J$ only controls the multiplet splitting at fixed charge, it is not affected by the approach 
to the Mott transition, hence $J\to J_*\simeq J$. On the contrary, the residual quasi-particle repulsion is
substantially weakened near a Mott transition, since most of correlation effects are already built into the small $Z$. 
On the basis of the DMFT behavior of the double occupancy in the single band Hubbard model,~\cite{DMFT} 
it was speculated in Ref.~\onlinecite{Capone-Science} that 
$U\to U_* \simeq Z\,U$. This assumption would result in the ansatz for the quasiparticle scattering amplitude 
in the Cooper channel (\ref{SC-chan}) 
\be
A = U -2J\rightarrow A_* = U_* -2J_* \simeq Z\,U - 2J.
\label{ansatz-A}
\ee
Should Eq. (\ref{ansatz-A}) be correct, it would imply that $A_*$, which for small $U>2J$ is repulsive in agreement with perturbation 
theory, must necessarily turn attractive sufficiently close to the Mott transition, where $Z$ goes to zero.
In Fig.~\ref{A*} we show $A_*$ given by Eq.~(\ref{ansatz-A}) with $Z$ extracted according to Eq.~(\ref{defZ}) from the normal-state 
DMFT solution. We note that this estimate for $A_*$ changes from repulsive to attractive very close to the point where a stable 
superconducting solution is found, supporting the validity of (\ref{ansatz-A}). Moreover, the expression (\ref{ansatz-A}) 
provides an explanation for the large strength of pairing in the SCS phase in comparison with the BCS state.
In fact, as $Z\to 0$, a regime in which $A_*\simeq -W_* = -Z\,W$ is eventually reached. Here the quasiparticles feel an attraction of 
the same order of their effective bandwidth. This intermediate regime, bridging between the weak-coupling BCS limit and the 
strong-coupling Bose regime, has been shown to be the optimal situation for superconductivity in purely attractive 
models\cite{Ranninger,rrattractive2}.

According to the above Fermi-liquid arguments, superconductivity re-emerges before the Mott transition because our Hamiltonian 
contains a pairing mechanism whose only role is to favor the singlet configuration (\ref{singlet}) whenever two electrons sit on the 
same site. For this reason pairing is strengthened rather than weakened as Mott localization is approached. A similar phenomenon 
is not expected to occur if pairing is mediated by a charge-charge attraction, that obviously conflicts with the Coulomb repulsion. 
Notice that in the present context the origin of $J$ (purely electronic generated, e.g., by superexchange, or driven by coupling 
with Jahn-Teller phonons as in fullerenes \cite{Gunnarsson-RMP}) is not important; the only relevant point being that it does not 
compete with $U$.

\section{Insights from the impurity model}

Although Fermi-liquid theory can be rather safely invoked to explain the emergence of superconductivity from the correlated metal, 
it has to be eventually abandoned in the region of coupling which precedes the Mott transition, where, as we discussed, 
superconductivity emerges out of a pseudogap normal phase. This demands for an alternative description able to account for both the 
Fermi-liquid and non-Fermi-liquid regimes.

As we introduced in Sec.~\ref{Sec:Model}, DMFT establishes a correspondence between a lattice model and an impurity model. 
The equivalence between the two models is enforced by a self-consistency condition which contains the information about the 
original lattice. 
The role of the DMFT self-consistency is absolutely non trivial: for instance it makes the impurity Kondo temperature $T_K$  
vanish at a finite value of the interaction which signals the Mott transition in the lattice model, since $T_K$ coincides 
with the renormalized bandwidth $W_*=Z\,W$. However, important insights can be gained by the analysis of the impurity model alone, 
without imposing the self-consistency. 

The Hamiltonian of the impurity model corresponding to (\ref{Ham}) is
\bea
\mathcal{H} &=& -\sum_{\bk}\,\sum_{a=1}^2\,\sum_\sigma\, \epsilon_\bk \, c^\dagger_{\bk,a\sigma}c^\dagga_{\bk,a\sigma} 
+ \Big(V_\bk\,c^\dagger_{\bk,a\sigma}d^\dagga_{a\sigma} + H.c.\Big) \nonumber \\
&& + \frac{U}{2}\left(n_d-2\right)^2 
-2J\, \Big( T_{x}^2 + T_{y}^2\Big),\label{Ham-AIM}
\eea 
where $c^\dagger_{\bk,a\sigma}$ and $c^\dagga_{\bk,a\sigma}$ are auxiliary fermionic degrees of freedom introduced to describe 
the Weiss field, while $d^\dagger_{a\sigma}$, $d^\dagga_{a\sigma}$, $n_d$ and $T_{x(y)(z)}$ are the same operators defined above, 
but they are specialized to the impurity site.
The physics of the model is controlled by three energy scales: $U$, $J$ and the so-called hybridization width $\Gamma$ defined through 
\be
\Gamma = \pi\sum_\bk\, \left|V_\bk\right|^2\,\delta\left(\epsilon_\bk\right).
\ee
The Hamiltonian (\ref{Ham-AIM}) has been recently studied in Refs.~\onlinecite{DeLeo-PRL-1,DeLeo04f,ferrero-2007} by means of 
Numerical Renormalization Group (NRG) for a constant hybridization function.~\cite{Krishnamurthy-1,Krishnamurthy-2} 
Analyzing the evolution of the physics as a function of $U$ at fixed $\Gamma$ and $J\ll \Gamma$, a quantum phase transition takes 
place at $U = U_*$, an interaction value that corresponds to a Kondo temperature $T_K^*\simeq J$. For $U<U_*$, i.e., $T_K > T_K^*$, 
perfect Kondo screening takes place. On the contrary, when $U>U_*$, $T_K<T_K^*$, the impurity locks into the singlet configuration 
\be
{\frac{1}{\sqrt{2}}}\,\left(d^\dagger_{1\su}d^\dagger_{2\giu}+d^\dagger_{2\su}d^\dagger_{1\giu}\right)\,|0\rangle,
\ee
which does not require any Kondo screening. The critical point that separates the two phases is similar to that found in the two 
impurity Kondo model~\cite{Jones89,Affleck95}. Moreover the critical point exists also away from particle-hole symmetry, and, 
remarkably, it can be accessed even for $U>U_*$ changing the density away from half-filling.~\cite{DeLeo04f}  
The single-particle spectral function displays an interesting evolution across the transition.~\cite{DeLeo04f} In the Kondo 
screened phase, the low-frequency DOS is characterized by a broad resonance around the Fermi level on top of which a narrow 
Kondo peak develops. The latter shrinks continuously as the critical point is approached, while the broad resonance stays unaffected 
as well as the high-energy Hubbard bands surrounding the low-energy features. At the critical point, the Kondo peak disappears 
leaving behind only the broad resonance. 
In the unscreened phase, a narrow pseudogap appears inside the broad resonance, whose width grows moving away from the critical point. 
This pseudogap is gradually filled away from particle-hole symmetry, although, as mentioned, the critical point still exists. 
This behavior has been parametrized~\cite{DeLeo04f} by the following ansatz for the 
low-energy DOS at particle-hole symmetry: 
\be
\rho(\epsilon) = \frac{1}{\pi\Gamma}\Big(\frac{T_+^2}{\epsilon^2+T_+^2}
\pm \frac{T_-^2}{\epsilon^2+T_-^2}\Big),
\label{ansatz-AIM}
\ee
where the plus and minus signs refer to the screened and unscreened phases, respectively. The energy scale $T_+\sim max(T_K,J)$ 
measures the width of the broad resonance, while $T_-\propto |U-U_*|^2$ controls the deviation from the fixed point and sets the 
width of the Kondo peak in the screened phase and of the pseudogap in the unscreened one.  
This model DOS defines a self-energy that was shown to fit perfectly well the numerical data for the impurity model.~\cite{DeLeo04f}  
In particular, the self-energy at small Matsubara frequencies is Fermi-liquid-like in the Kondo screened phase, 
$\Sigma(i \omega_n)\propto -i\omega_n$, is finite and imaginary at the critical point, $\Sigma(i\omega_n)=-i\Gamma$, 
and diverges in the unscreened phase, $\Sigma(i\omega_n) \propto 1/i\omega_n$. 

The critical point is unstable in several symmetry breaking channels~\cite{Affleck95,DeLeo-PRL-1}, 
including the magnetic channel 
\be
\label{magn-AF}
\mathbf{S}_1 - \mathbf{S}_2,
\ee
the Cooper channel Eq.~(\ref{SC-chan}), and the hybridization channels 
\be
\label{Hyb-chan}
 \sum_\sigma\,d^\dagger_{1\sigma}d^\dagga_{2\sigma},\;  \sum_\sigma\,d^\dagger_{2\sigma}d^\dagga_{1\sigma},
\ee
that break the $O(2)$ orbital symmetry.  
All these channels are degenerate at the critical point, where the model has an enlarged $SO(7)$ symmetry.~\cite{Affleck95}
When the impurity model contains explicitly a symmetry-breaking term that couples to one of the unstable channels, the critical point 
is washed out and the phase transition between the Kondo and the local singlet phase turns into a 
sharp cross-over.~\cite{DeLeo04f,ferrero-2007} In the meantime, the relevant perturbation cuts off the non-Fermi-liquid singularities 
of the self-energy both at the critical point and in the unscreened phase, so that, at sufficiently low frequency, 
$\Sigma(i \omega_n)\propto -i\omega_n$ is always recovered.~\cite{ferrero-2007}  

Since the Kondo temperature $T_K$ has to vanish as the Mott transition is approached, the effective impurity model must encounter 
the critical point, $T_K\simeq J$, before the Mott point is reached.
In Ref.~\onlinecite{DeLeo-PRL-1} it has been speculated that, once a full DMFT calculation is carried on, the instabilities 
associated to the impurity critical point lead because of the self-consistency condition to a spontaneous symmetry breaking in a 
whole region around the critical point along one of the instability channels. These are the particle-hole channels (\ref{magn-AF}) 
and (\ref{Hyb-chan}), which correspond to magnetic and orbital ordering, respectively, and the particle-particle channel 
(\ref{SC-chan}) that implies superconductivity.
This prediction is perfectly compatible with the DMFT phase diagram, Fig.~\ref{phd}.
We note that particle-hole instabilities usually require nesting or other band-structure singularities that only 
accidentally occur, while the Cooper singularity is more ubiquitous. For this reason, in Ref.~\onlinecite{Capone-PRL-2} 
we only searched for a superconducting instability, even though the Bethe lattice with nearest neighbor hopping 
has nesting at half-filling.

\section{Modeling the dynamics in the normal phases}

\begin{figure}[htb]
\includegraphics[width=8.cm]{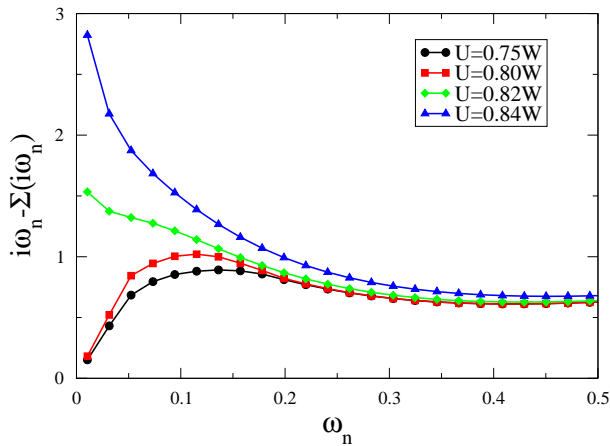}
\caption{\label{Sigma_N} (Color online) $i\omega_n-\Sigma(i\omega_n)$ versus $\omega_n>0$ for different values of $U/W$, 
as obtained by DMFT preventing superconductivity.}
\end{figure} 
The analysis of the previous section shows that the critical point of the impurity model is expected to be always pre-empted by 
broken symmetry phases in the lattice model treated within DMFT. Nonetheless, it is plausible that the critical point correspond to a 
metastable phase, just like a normal metal is a metastable phase in the presence of attraction. 
Following this idea, we have solved the model (\ref{Ham}) forcing the DMFT self-consistency not to break any symmetry. 
The behavior of the self-energy that we obtain, see Fig.~\ref{Sigma_N}, closely reminds that of the impurity model with constant 
bath (i.e. without self-consistency) that we just discussed.
This confirms that the impurity critical point corresponds to a metastable phase in the lattice model and suggests that the same 
parametrization Eq.~(\ref{ansatz-AIM}) may work in the lattice as well.
Therefore we have assumed for the low frequency local Green's function the following expression, valid in a Bethe lattice 
(all energies will be expressed in units of the bandwidth $W=1$),  
\be
\mathcal{G}(i\omega_n) = \frac{1}{2}\left[ \mathcal{G}_0\left(\frac{i\omega_n}{T_+}\right)
\pm  \mathcal{G}_0\left(\frac{i\omega_n}{T_-}\right)\right],
\label{ansatz-DMFT}
\ee
where  the $\pm$ sign, $T_+$ and $T_-$ have the same meaning as in Eq.~(\ref{ansatz-AIM}), while 
\be
\mathcal{G}_0(i\omega_n) = 8i\left[\omega_n -\mathrm{sign}(\omega_n)\, \sqrt{\omega_n^2 + \frac{1}{4}}\right],
\ee
is the non-interacting local Green's function. The ansatz (\ref{ansatz-DMFT}) for the Green's function 
corresponds to an ansatz for the self-energy which, through the 
DMFT self-consistency equation, becomes 
\be
\Sigma(i\omega_n) = i\omega_n - \frac{1}{16}\mathcal{G}(i\omega_n) - \mathcal{G}(i\omega_n)^{-1}. 
\label{ansatz-DMFT-Sigma}
\ee
In the Fermi liquid region, where the plus sign has to be used in (\ref{ansatz-DMFT})and $T_-\not = 0$, the low-frequency self-energy is  
\bea
\Sigma(i\omega_n) &\simeq& i\omega_n -i\,\frac{\omega_n}{2}\left(\frac{1}{T_+}+\frac{1}{T_-}\right)\nonumber \\
&& + i\,\frac{\omega_n^2}{4}\left(\frac{1}{T_+}-\frac{1}{T_-}\right)^2\,\mathrm{sign}(\omega_n),
\label{Sigma+}
\eea
corresponding to a regular Fermi liquid behavior with quasi-particle residue
\be
Z = 2\,\frac{T_+\,T_-}{T_++T_-}.
\label{Z-ansatz}
\ee
At the critical point ($T_-=0$), or for frequencies $T_-\ll\omega_n\ll T_+$, the self-energy becomes
\bea
\Sigma(i\omega_n) &\simeq& i\omega_n -i\,\frac{3}{8}\,\mathrm{sign}(\omega_n) 
- i\,\frac{5\omega_n}{4T_+}\nonumber \\
&& -i\,\frac{3}{4}\left(\frac{\omega_n}{T_+}\right)^2\,\mathrm{sign}(\omega_n).
\label{Sigma*}
\eea
$\Sigma(i\omega_n)$ has a finite and sizeable imaginary part, implying a non-Fermi liquid behavior. 
The deviations from conventional Fermi liquid behavior become even more pronounced in the unscreened phase, 
where $T_-\not=0$ and the minus sign has to be used in (\ref{ansatz-DMFT}). Here we obtain
\bea
\Sigma(i\omega_n) &\simeq& i\omega_n - \frac{i}{4\omega_n}\,\frac{T_+\,T_-}{T_+-T_-}\nonumber\\
&&- \frac{i}{4}\,\frac{T_++T_-}{T_+-T_-}\,\mathrm{sign}(\omega_n) - i\,\frac{\omega_n}{T_+-T_-}.
\label{Sigma-}
\eea
The divergence of $\Sigma\left(i\omega_n\right)$ for small $\omega_n$ leads to a pseudo-gap in the DOS, 
whose behavior at small energy is 
\be
\rho(\epsilon) \simeq 4\,\left(\frac{1}{T_-^2}-\frac{1}{T_+^2}\right)\,\epsilon^2.
\ee

\begin{figure}[htb]
\includegraphics[width=8.cm]{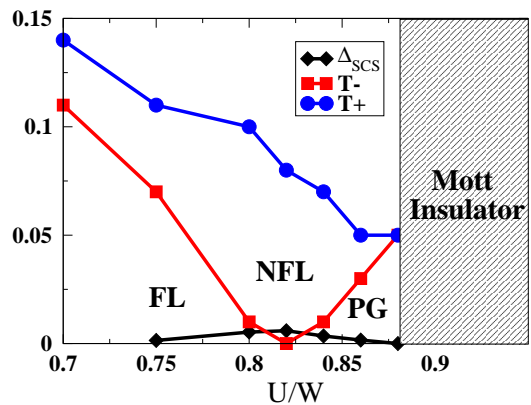}
\caption{\label{scale} (Color online) Fitting parameters $T_+$ and $T_-$ of the DMFT normal self-energy with the expression Eq.~({ansatz-DMFT-Sigma}) 
using for the impurity Green's function Eq.~(\ref{ansatz-DMFT}). We also draw the DMFT superconducting 
gap when gauge symmetry breaking is allowed.}
\end{figure}

We can check the validity of our ansatz by simply comparing it with the actual DMFT results obtained using exact diagonalization. 
Fitting these data with (\ref{ansatz-DMFT}) we find a very good agreement, and we can extract the parameters $T_+$ and $T_-$. 
The behavior of the best-fit values of $T_+$ and $T_-$ is drawn in Fig. \ref{scale} and supports our ansatz, and consequently the 
relevance of the impurity critical point for the lattice model. In the same figure we report the superconducting gap obtained 
in DMFT allowing for gauge-symmetry breaking.

Two things are worth noting in this figure. First, the vanishing of $T_-$ at $U\simeq 0.82~W$, namely the location of the  
{\sl metastable} critical point, corresponds to the maximum of the superconducting gap once gauge symmetry breaking is allowed, 
in accordance with the prediction based on the analysis of the impurity model.~\cite{DeLeo-PRL-1} 
At the metastable critical point, the only remaining energy scale is $T_+$ that therefore determines the maximum value of the 
superconducting gap $\Delta$.
Secondly, the Mott transition occurs within the pseudogap region when $T_-=T_+$. In other words, the pseudogap gradually widens as 
$U$ increases from the $T_-=0$ point and, at the same time, the total spectral weight of the low-energy part, given by $(T_+-T_-)$ 
according to Eq.~(\ref{ansatz-DMFT}), diminishes until it disappears, when $T_-=T_+$.

Although we solved DMFT at zero temperature, we believe that the $T=0$ energy scales will at least approximately determine the 
finite-temperature behavior of the model. We can therefore speculate that the energy scales $T_+$, $T_-$ and $\Delta$ will reflect 
in analogous temperature scales, leading to the scenario drawn in Fig.~\ref{phd-rough} and discussed in the Introduction. 
In this perspective, even though the critical region around $T_-=0$ is not stable at $T=0$, and in fact in our calculation is replaced 
by superconductivity, it should become accessible by raising temperature $T$ when $T_+>T>Max(T_-,T_c)$, denoted as NFL 
(non-Fermi-liquid) in the figure. If we follow a path in the $T-U$ space that avoids the bell-shaped superconducting region, we must 
find a crossover from a correlated Fermi-liquid phase to a pseudogap state passing through the NFL region.

Finally, the region above $T_+$ has a simple interpretation in the impurity model, where it corresponds to the local moment regime 
above the Kondo temperature. In the lattice model, it presumably translates into a phase with very poor lattice-coherence, although 
its precise properties are difficult to foresee in the absence of actual finite-$T$ calculations.

\section{Modeling the dynamics in the superconducting phase}
\label{secdisorder}

An important result for the impurity model is that, as soon as one of the relevant symmetry breaking perturbation is introduced, 
like, e.g., the hybridization (\ref{Hyb-chan}), suddenly the non-Fermi liquid behavior of the self-energy is replaced by a standard 
Fermi liquid one. In this case the low-frequency self-energy at the critical point changes from an imaginary constant to a conventional 
linearly vanishing function.~\cite{ferrero-2007}. 
We can therefore expect a similar regularization to occur as soon as superconductivity is allowed in the lattice model. 
This expectation is closely reminiscent of the mechanism taking place when $s$-wave superconductivity establishes 
in a disordered metal. Here scattering off impurities makes the quasiparticle lifetime $1/\tau$ finite, 
which means that the normal state self-energy is finite and imaginary at zero frequency, just like in the above 
discussed non-Fermi liquid phase. Following Abrikosov, Gor'kov and Dzyaloshinskii,~\cite{abrikosov}, 
we write the self-energy as a 2$\times$2 matrix whose diagonal entry $\Sigma_{11}$ is the normal component, and 
the off-diagonal $\Sigma_{12}$ is the anomalous (superconducting) contribution. 
We define the function $\eta(i\omega_n)$ in the normal phase according to
\be
i\omega_n - \Sigma_{11}(i\omega_n) = i\omega_n + \frac{i}{2\tau}\,\mathrm{sign}(\omega_n) \equiv i\omega_n\,\eta(i\omega_n).
\label{disorder}
\ee
The onset of superconductivity regularizes this normal-state anomaly giving rise to a normal self-energy $\Sigma_{11}$ linear below 
a low-energy scale $\Delta$ and, at the same time, the anomalous self-energy $\Sigma_{12}$ gets strongly enhanced with respect 
to a clean superconductor, namely~\cite{abrikosov}
\begin{eqnarray}
\label{sigma11-dis}
\io-\Sigma_{11}\left(\io\right) &=&\io\,\eta\left(i\sqrt{\omega_n^2+\Delta^2}\right)\\
\label{sigma12-dis}
\Sigma_{12}\left(\io\right)&=&\Delta\,\eta\left(i\sqrt{\omega_n^2+\Delta^2}\right), 
\end{eqnarray}
$\eta$ being the same function defined in Eq. (\ref{disorder}).
The two above equations express an important physical property, namely that non-magnetic disorder is a non pairbreaking perturbation 
in a BCS superconductor. Indeed Eqs.~(\ref{sigma11-dis}) and (\ref{sigma12-dis}) imply a perfect cancellation in the $s$-wave 
Cooper channel between the wave-function renormalization (the self-energy) and the vertex corrections brought by the impurities. 
This leads to the well-known ``Anderson's theorem'',~\cite{anderson-theorem} stating that the value of $T_c$ is independent of 
the concentration of non magnetic impurities, provided the latter is low. This result follows immediately from the BCS gap-equation 
in the presence of an attractive coupling $\lambda$~\cite{abrikosov}
\be
1 = \lambda\,T\,\sum_{\io}\,\sum_\bk\, \frac{\eta\left(i\sqrt{\omega_n^2+\Delta^2}\right)}
{\left(\omega_n^2+\Delta^2\right)\,\eta(i\sqrt{\omega_n^2+\Delta^2})+\epsilon_\bk^2}.
\label{BCS}
\ee
$T_c$ is determined by solving (\ref{BCS}) with $\Delta=0$, namely $\eta(\io) = 1 + 1/(2\tau|\omega_n|)$.  One readily 
realizes that, summing over momentum first, $\eta(\io)$ disappears from the equation in the infinite bandwidth limit, leading to 
the same logarithmically singular sum over Matsubara frequencies as in the absence of disorder.

We discussed previously that the emergence of SCS can be explained assuming that vertex corrections compensate exactly for 
the strong wave-function renormalization. It is then tempting to further pursue the analogy with dirty $s$-wave superconductors. 
Namely, we define
\be
\io - \Sigma_{11}(\io) \equiv \frac{\displaystyle \io}{\displaystyle Z(\io)},
\label{def}
\ee
where $\Sigma_{11}(\io)$ is the self-energy of the metastable normal solution, using instead of $\eta(\io)$, the more conventional 
notation $Z(\io)$ for the frequency dependent wave-function renormalization. 
We then assume that allowing for superconductivity leads to diagonal, $\Sigma_{11}$, and off-diagonal, $\Sigma_{12}$, self-energy matrix elements, in the Nambu-spinor notation, given as in (\ref{sigma11-dis}) and (\ref{sigma12-dis}) by  
\begin{eqnarray}
\label{sigma11}
\io-\Sigma_{11}\left(\io\right) &=&\frac{\displaystyle \io}{\displaystyle Z\left(i\sqrt{\omega_n^2+\Delta^2}\right)}\\
\label{sigma12}
\Sigma_{12}\left(\io\right)&=& \frac{\displaystyle \Delta}{\displaystyle Z\left(i\sqrt{\omega_n^2+\Delta^2}\right)},
\end{eqnarray} 
where $\Delta$ is the cut-off scale introduced by superconductivity. It follows immediately that the introduction of 
$\Delta$ restores a conventional Fermi-liquid behavior, $\Sigma_{11}\sim -\io$, not only when $T_-=0$ and $Z(\io)\sim |\omega_n|$, 
but also in the pseudogap regime where $Z(\io)\sim |\omega_n|^2$. 
Within this picture superconductivity regularizes the low-frequency behavior of the normal state self-energy both at the 
critical point, which is actually avoided by the onset of symmetry breaking, and in pseudogapped phase whose singularity is 
cut-off by $\Delta$. As a consequence of such regularization, the low-energy pseudogap in the spectral function should change 
within the superconducting phase into rather sharp quasi-particle peaks at the edge of the gap which finally disappears 
approaching the Mott insulator.
\begin{figure}[htb]
\includegraphics[width=9.5cm]{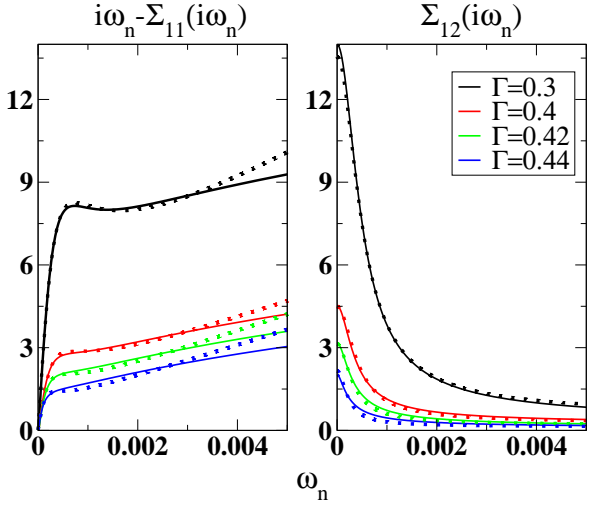}
\caption{\label{sigma} (Color online) Fit of the impurity diagonal and off-diagonal self-energy, in the Nambu-spinor space, with 
the expressions Eqs.~(\ref{sigma11}) and (\ref{sigma12}).}
\end{figure}

Unfortunately a thorough comparison of numerical DMFT data with (\ref{sigma11}) and (\ref{sigma12}) is a very hard task, 
because the phenomena we want to observe involve the extremely small-frequency range (where the Fermi-liquid behavior would be 
recovered), which is really hard to study with present impurity solvers. 
For this reason, since the idea we propose is quite general, we have decided to verify the validity of our ansatz for the impurity 
model without self-consistency, postponing some selected DMFT comparison to the end of the section.

\begin{figure}[htb]
\includegraphics[width=8cm]{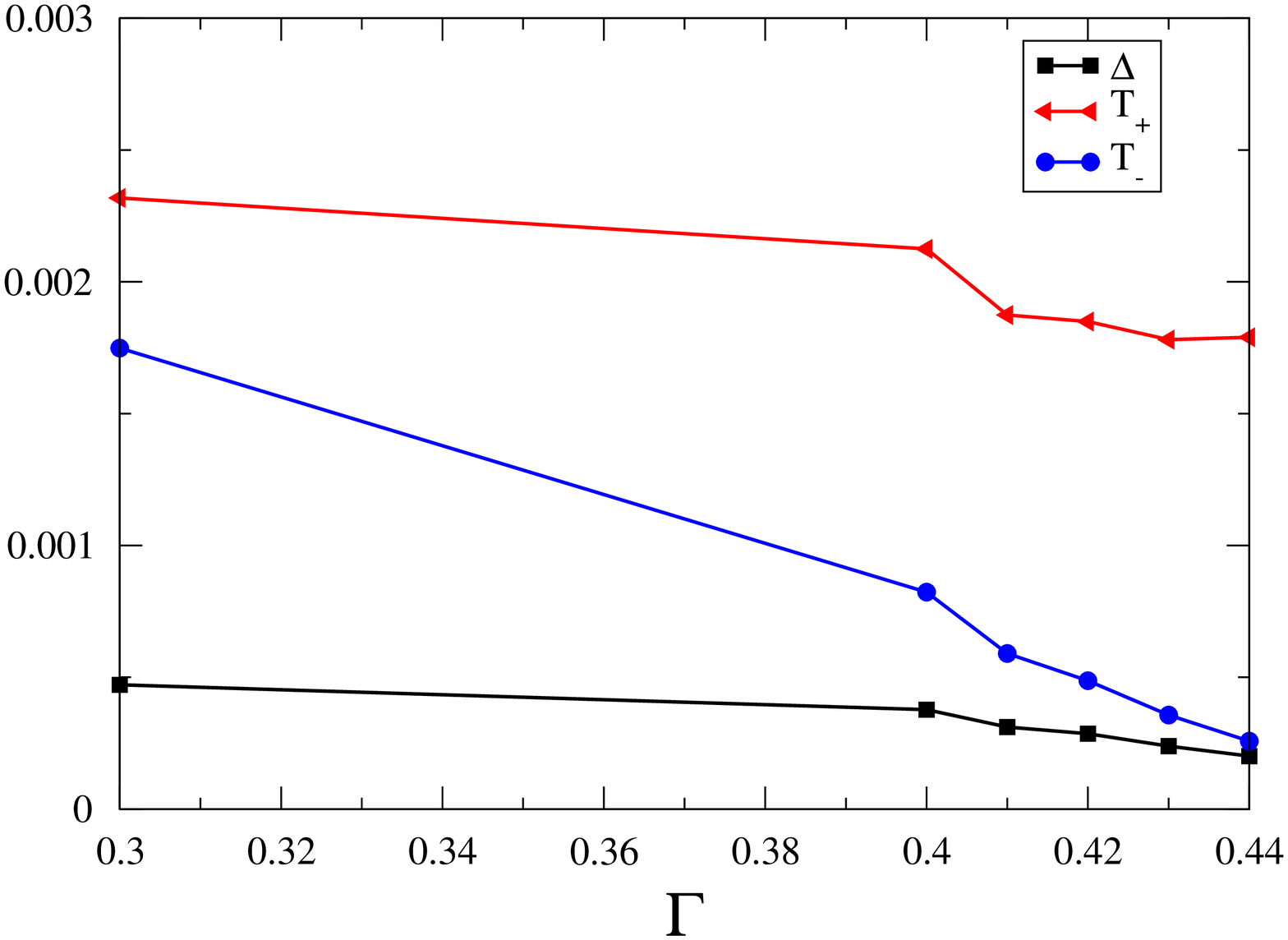}
\caption{\label{fit-parameter} (Color online) Fitting parameters $T_+$, $T_-$ and $\Delta$ of the impurity-model self-energy with the ansatz Eqs.~(\ref{sigma11}) and (\ref{sigma12}) as a function of the hybridization $\Gamma$.}
\end{figure}

\begin{figure}[hbt]
\includegraphics[width=8.cm]{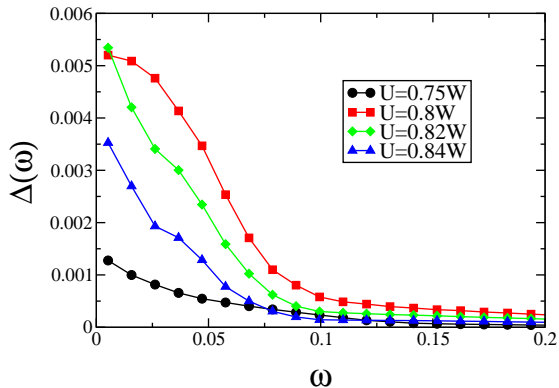}
\caption{\label{s} (Color online) Frequency-dependent superconducting gap for several values of $U$.}
\end{figure}

An important difference is that no spontaneous symmetry breaking is possible for the impurity model without self-consistency. 
Therefore, in order to check the validity of (\ref{sigma11}) and (\ref{sigma12}), one has to add explicitly a 
symmetry-breaking perturbation to the Hamiltonian (\ref{Ham-AIM}). In Refs.~\onlinecite{DeLeo04f,ferrero-2007} 
the impurity model
\bea
\mathcal{H} &=& -\sum_{\bk}\,\sum_{a=1}^2\,\sum_\sigma\, \epsilon_\bk \, c^\dagger_{\bk,a\sigma}c^\dagga_{\bk,a\sigma} 
+ \Big(V_\bk\,c^\dagger_{\bk,a\sigma}d^\dagga_{a\sigma} + H.c.\Big) \nonumber \\
&& -t_\perp\sum_\sigma \,\Big(d^\dagger_{1\sigma}d^\dagga_{2\sigma} + H.c.\Big),\label{Ham-AIM-2}
\eea 
was considered with $U=8$, $t_\perp=0.05$ and different $\Gamma$'s, in units of half the conduction bandwidth. 
In this model the role of the hybridization $t_\perp$ is twofold. On one hand it 
generates an exchange $J_\perp=4t_\perp^2/U$ able to drive the model through the critical point. At the same time, 
$t_\perp$ breaks the $O(2)$ orbital symmetry, turning the quantum phase transition into a crossover which has been shown to be quite 
sharp.~\cite{DeLeo04f,ferrero-2007} We note that, at particle-hole symmetry, the two channels (\ref{SC-chan}) and (\ref{Hyb-chan}) 
are perfectly equivalent, being related by the particle-hole transformation
\bea
d^\dagga_{2\su} \to d^\dagger_{2\giu} \qquad & d^\dagga_{2\giu} \to -d^\dagger_{2\su}\\
c^\dagga_{\bk 2\su} \to -c^\dagger_{\bk_* 2\giu} \qquad & c^\dagga_{\bk 2\giu} \to c^\dagger_{\bk_*2\su}\\
\eea
where $\bk$ and $\bk_*$ are particle-hole partners, $\epsilon_\bk=-\epsilon_{\bk_*}$ and $V_{\bk_*} = V_\bk^*$.  
Therefore we can simply borrow the NRG data of Refs.~\onlinecite{DeLeo04f,ferrero-2007} and adapt them to our case of a 
superconducting symmetry-breaking term. In Fig.~\ref{sigma} we show the NRG data against our fit using Eqs.~(\ref{sigma11}) 
and (\ref{sigma12}) with the normal self-energy that follows from the model DOS (\ref{ansatz-AIM}). 
The fitting parameters $T_+$, $T_-$ and $\Delta$ are shown as function of $\Gamma$ in Fig.~\ref{fit-parameter}.~\cite{nota} 
We take the validity of Eqs. (\ref{sigma11}) and (\ref{sigma12}) for the impurity model as a strong support of their validity 
also in the lattice model even deep inside the pseudogap phase where, as previously mentioned, our numerical data have not enough 
precision to reveal the very low-frequency structure.
An interesting feature of the off-diagonal self-energy in the impurity model is the strong frequency dependence, especially in the 
pseudogap region where $\Sigma_{12}$ is extremely peaked at very low frequencies.
This is true also in the actual DMFT calculation. In Fig.~\ref{s} we plot, for different $U$'s 
around the top of the SCS region, $U\simeq 0.82~W$, the frequency dependent 
superconducting gap, defined by
\be
\Delta(i\omega_n) = \frac{\displaystyle i\omega_n\, \Sigma_{12}(i\omega_n)}
{\displaystyle i\omega_n - \Sigma_{11}(i\omega_n)},
\label{def:Delta}
\ee
whose zero-frequency extrapolation is the gap shown in Fig.~\ref{gap}. We note the rapid rise of $\Delta(i\omega_n)$ 
below a frequency of order $T_+$, confirming our prediction that, around its maximum, the gap is controlled 
by a single energy scale, $T_+$. Therefore, even though the attraction $J$ is instantaneous, 
strong retardation effects develop to avoid the large repulsion $U$ and stabilize superconductivity. 
It is just this strong frequency dependence of the superconducting gap (\ref{def:Delta}) that marks the difference 
between a conventional Bose-Einstein condensation of preformed local pairs and our SCS phase, where  
pairs are highly non-local in time, see Fig.~\ref{s}. We also note that a characteristic energy scale shows up in the 
dynamical gap function even if our model does not involve any exchange of bosons with a typical energy scale 
(at least not in an obvious way), as it has been instead proposed for the two-dimensional Hubbard model\cite{scalapino}.

Finally, we conclude this section by discussing the possibility to write down a gap-equation like Eq. (\ref{BCS}) 
for our model in the SCS phase, which remains an open and suggestive issue. Starting from the Fermi-liquid side, 
our DMFT results seem to suggest that pairing essentially involves only the strongly renormalized quasiparticles. 
In this case a natural candidate for a gap equation would be
\begin{equation}\label{eqn:gap_eq}
\Delta=A_*\,T\sum_{n}\sum_{\0{k}}F_{qp}\left(\io,\0{k}\right)\,,
\end{equation}
where $A_*$ is the renormalized scattering amplitude in the Cooper channel defined through Eq.~(\ref{ansatz-A}) while $F_{qp}$ 
is the quasiparticle anomalous Green's function
\begin{equation}
\label{fqp}
F_{qp}=\frac{1}{Z}\,F\left(\io,\epsilon_{\bk}\right)\,.
\end{equation}
A first analysis shows that Eq.~(\ref{eqn:gap_eq}) correctly reproduces the order of magnitude of $\Delta$ 
also in the strongly correlated regime. This encourages us to use it to estimate the effective attraction $A_*$ 
such that Eq.~(\ref{eqn:gap_eq}) gives the actual DMFT value of $\Delta$. We use (\ref{eqn:gap_eq}) and (\ref{fqp}) also in the 
pseudogap state, where the Fermi-liquid ansatz for $A_*$, Eq.~(\ref{ansatz-A}), is not valid. Here $Z$ is taken to be 
the low-energy spectral weight within a window of the order $4J$ around the Fermi level.
\begin{figure}[htb]
\includegraphics[width=8.cm]{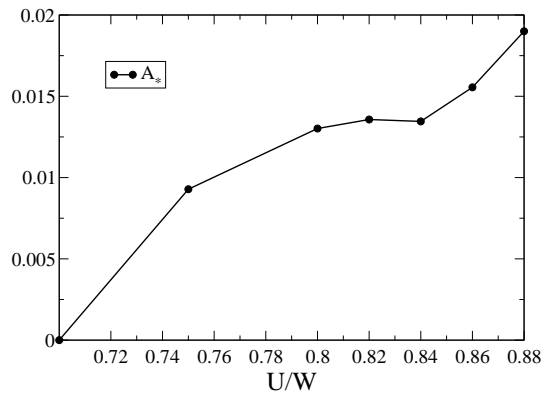}
\caption{\label{lambda} Attractive coupling constant $A_*$ extracted through Eq.~(\ref{BCS}) as the value
that gives the correct DMFT value of $\Delta$.}
\end{figure} 
$A_*$ that we obtain this way, shown in Fig.~\ref{lambda}, is quite smooth as a function of $U$ and it stays relatively small, 
in the range between 0.01 and 0.02, 
up to the Mott transition. This shows that, even in the pseudogap regime, Eq.~(\ref{eqn:gap_eq}) can be satisfied without 
requiring a big coupling constant.

\section{Discussion and conclusions}

It is common wisdom that the emergence of local moments out of the incipient Mott localization must necessarily 
bring to an enhanced magnetic response and eventually to a magnetic order that may appear already in the metallic phase adjacent 
the Mott insulator. It is equally conceivable that, under these circumstances, a system may become easily unstable to lattice 
distortions.  Moreover, on the brink of charge localization, also the orbital momentum, quenched by the hopping deep in the 
metallic phase, may re-emerge making spin-orbit coupling effective as if atoms were isolated. This is the typical 
phenomenology of magnetic Mott insulators exemplified by the prototype Mott-Hubbard system, the V$_2$O$_3$.~\cite{Bao-PRB,Paolasini}
These properties can be interpreted within a Landau Fermi-Liquid framework only by invoking an almost perfect cancellation 
between the large wave function renormalization associated to the Mott transition 
and the vertex corrections in peculiar channels. Physical intuition suggests that such channels should be primarily those acting 
on degrees of freedom orthogonal to charge, like spin and orbital momentum. 

However, while the above arguments are quite natural, if not obvious, for particle-hole instabilities like magnetism, 
they sound much less trivial in connection with superconducting particle-particle instabilities. The main difference is 
that superconductivity implies phase-coherence, which can not survive charge localization. Nevertheless, while phase-coherence 
requires pairing, the opposite is not true, since a Mott insulator can be formed by incoherent pairs.
An example is a valence-bond crystal formed by an ordered array of tightly bound singlets sitting on nearest-neighbor bonds, 
and, seemingly, also a Mott insulator made of resonating valence bonds - the RVB scenario originally proposed by Anderson~\cite{PWA} 
for high-$T_c$ superconductors. 

Another possibility is to form local singlets exploiting the orbital degrees of freedom, 
that play a role similar to the bonds of the previous examples. This is realized in our model (\ref{Ham}), and in model for fullerenes.
In the fullerene family, tetravalent alkali-doped C$_{60}$ may be regarded as the parent Mott insulating 
compound~\cite{Capone-PRL-1,Capone-Science} that turns superconducting upon doping, the trivalent materials. The former compounds 
are indeed non-magnetic Mott insulators where the four electrons occupying the LUMO of each molecule bind into a non degenerate 
spin-singlet configuration because of Jahn-Teller effect.~\cite{Michele-PRB-Mott-JT} 

In all these examples, the pairing implicit in the Mott state is perfectly compatible with strong repulsion, hence it is not 
surprising that, moving away from the Mott insulator, superconductivity appears. 
The general conditions are elucidated by our analysis: pairing must correlate degrees of freedom orthogonal to charge.
For instance, in the $t$-$J$ model for cuprates pairing is provided by the spin superexchange, that is unaffected by the strong 
repulsion constraint of no-double occupancy, while it competes with the hopping. The former favors configurations in which 
two singly-occupied nearest neighbor sites are bound into a singlet state, while the latter prefers a democratic occupancy, 
in which the singlet is equally probable as any of the triplet states.       
Also in our two-orbital model (\ref{Ham}) close to the Mott transition, $J$ competes with the hopping rather than $U$. 
At half-filling the hopping favors an equal occupation of all the states with two electrons per site, while $J$ breaks this 
degeneracy in favor of  the singlet configuration (\ref{singlet}). Right the same competition emerges in the impurity 
model (\ref{Ham-AIM}) between the Kondo temperature and $J$. The reason why $J\ll t$ eventually prevails in all the above 
examples is that the hopping suffers from a very severe wavefunction renormalization close to a Mott transition, while $J$ does not. 

From these observations it appears evident that the physics of our model (\ref{Ham}) and that of the $t$-$J$ model for cuprates 
within the RVB scenario~\cite{PWA,Kotliar&Liu,Suzumura-RVB,Lee&Nagaosa&Wen-RMP} share common features; our inverted exchange 
playing on-site a similar role as the nearest-neighbor exchange in the $t$-$J$ model. Quite obviously $J$ being on-site or on a link 
is expected to introduce relevant differences as far as the momentum structure is concerned (the most evident being $s$-wave 
pairing versus $d$-wave pairing). 
There are however other relevant differences, at least regarding the interpretation of the various phases, 
In the RVB scenario uncovered by slave-boson~\cite{Kotliar&Liu,Baskaran,Suzumura-RVB,Lee&Nagaosa&Wen-RMP} and 
variational~\cite{Gros,Sorella&Dagotto,Paramekanti,Vanilla,Ivanov} approaches, a lot of emphasis is placed on the 
pseudogap regime close to the Mott transition to explain superconductivity.~\cite{Lee&Nagaosa&Wen-RMP} 
In our model superconductivity is instead the low-temperature response to the instability in the crossover region between 
the pseudogap and the Fermi-liquid. Another important difference is that our pseudogap state is a stable thermodynamic phase 
with zero entropy at zero temperature, while in the RVB scenario it appears as a fluctuating mixture of competing orders, 
unable to survive down to zero temperature.

Our results look also different from QCP theories for the cuprates~\cite{Varma-QCP,Castellani-QCP,Tallon&Loram-QCP}. 
Within the QCP approach superconductivity arises because of the critical fluctuations around a true quantum critical point that 
separates stable zero-temperature phases. This critical point may even be inaccessible in the physical space of parameters, 
and exist only in an hypothetical enlarged space, yet it is believed to influence the physical system not only at finite but also 
down to zero temperature. In contrast, in model (\ref{Ham}) the normal Fermi-liquid metal, the pseudogap phase, the superconductor 
and other possible symmetry-broken phases are all equally legitimate outcomes of an underlying competition that reveals itself only in 
the high-temperature non-Fermi-liquid crossover region, NFL in Fig.~\ref{scale}. From this point of view, the difference between the 
phase diagram Fig.~\ref{scale} and the QCP scenario is the same emphasized recently by Anderson.~\cite{PWA-QCP} An obvious and 
pertinent criticism to our claim is that, upon applying a sufficiently high magnetic field able to completely wash out 
superconductivity, the unstable critical region should move down to zero temperature and transform into a true quantum critical point 
thus recovering the QCP scenario. We note however that, at its maximum, the superconducting gap is only controlled by the energy scale 
$T_+$, which is also the bandwidth of the low-energy incoherent excitations. For this reason we believe, although we can not prove it, 
that a magnetic field so strong to completely suppress superconductivity would drive the model into a phase with very poor lattice 
coherence, rather then revealing the critical point. 

We believe that our results provide a new perspective about the relation between a pseudogap and superconductivity, which crucially 
differs from previous approaches. Rather than being in competition, the pseudogap and the superconducting gap 
turn out to be much more ``compatible'' one another than in most theoretical approaches that we are aware of. 
A central aspect, discussed in Sec.~\ref{secdisorder}, is that the large normal self-energy responsible of the pseudogap state 
is regularized by the onset of superconductivity, in contrast with approaches in which the same (highly anomalous) normal 
self-energy is assumed in the normal and superconducting states\cite{Rice-PG}. Moreover, we have shown that in model (\ref{Ham}) 
there is no need of a large coupling constant to overcome the lack of low-energy density of states in the normal phase, 
as it happens instead in approaches where pairing takes place starting from a normal state with mean-field like 
pseudogaps.\cite{nozieres-pistolesi,benfatto-caprara-dicastro}.

Another important outcome of our calculation is the natural appearance of two energy gaps with different behavior as a function of 
the distance from the Mott insulator (see Fig.~\ref{phd-rough}): the pseudogap scale  $T_-$ increases approaching the Mott state, 
where it is largest, while the superconducting coherence scale vanishes as the Mott insulator is approached. This is clearly 
reminiscent of the two energy scales (gaps) observed in the cuprates\cite{twogap,twogap2}. 

Finally, we want to emphasize that the physics we have unraveled seems to be more general than the specific toy-model (\ref{Ham}). 
Indeed, we have shown that the emergence of a reinforced superconductivity upon approaching the Mott transition can be explained 
in the model (\ref{Ham}) by simple Fermi-liquid arguments that do not depend on the specific model. 
They simply state that any scattering channel involving degrees of freedom orthogonal to charge should 
strengthen near a Mott transition, irrespectively whether being particle-particle or a particle-hole channels. 
This is for instance obvious in one dimension because of the dynamical separation of the charge from the other degrees of freedom, 
and explains why the phase diagram of (\ref{Ham}) in one dimension does not differ qualitatively from 
Fig.~\ref{scale}.~\cite{Fabrizio-1D-PRL} In addition, the impurity-model analogy, that we have exploited in this work, 
indicates that some kind of pseudogap phase - bulk counterpart of an impurity unscreened regime - 
might be ubiquitous near a Mott transition, especially if this transition is continuous. Indeed it is obvious that a metal 
should suppress spectral weight around the chemical potential to smoothly connect to a realistic zero-entropy insulator, and, 
in the absence of symmetry breaking, we do not see any other way but opening a pseudogap. 
This would in turn imply also the existence of a high-temperature cross-over regime that separates the Fermi-liquid phase from the 
pseudogap one and actually contains the seeds of the symmetry broken phase that eventually emerge at low 
temperature.~\cite{ferrero-2007} However, it often happens that the Mott transition is first order, in particular when accompanied 
by a lattice distortion, like in V$_2$O$_3$.~\cite{Bao-PRB} In this case both the pseudogap and the critical cross-over regions 
might become unaccessible and a discontinuous transition occur directly from the Fermi-liquid metal to the Mott insulator.

\begin{acknowledgments}
We are grateful to Lorenzo De Leo for providing us with his numerical renormalization group data on the impurity model. 
\end{acknowledgments}


\begin{thebibliography}{59}
\bibitem[{not()}]{nota}
\bibinfo{note}{Since the impurity model contains explicitly the symmetry
  breaking term, the self-energy that we fit is actually the off-diagonal
  component of the inverse Green's function.}
\end{thebibliography}

\end{document}